\documentclass[final,3p,times]{elsarticle}

\usepackage[singlelinecheck=false]{caption}
\usepackage{pgf,pgfarrows,pgfnodes,pgfautomata,pgfheaps}
\usepackage{graphicx}
\usepackage{subfigure}
\usepackage[T1]{fontenc}
\usepackage{amsfonts}
\usepackage{natbib}
\usepackage{url}
\usepackage{amsmath}
\usepackage{enumerate}
\usepackage{algorithmic}
\usepackage{algorithm}
\usepackage{amsmath}
\usepackage{amsfonts}
\usepackage{amssymb}

\usepackage{marvosym}
\usepackage{color}
\usepackage{diagbox}
\usepackage{tikz}
\usepackage{nomencl}
\usepackage{nomencl}
\usepackage{float}

\usepackage{multirow}
\usepackage{bm}
\def\BibTeX{{\rm B\kern-.05em{\sc i\kern-.025em b}\kern-.08em
		T\kern-.1667em\lower.7ex\hbox{E}\kern-.125emX}}	
\makeatletter

\newcommand{\Rmnum}[1]{\expandafter\@slowromancap\romannumeral #1@}
\makeatother
\usepackage[hidelinks]{hyperref}
\usepackage{caption}
\usepackage{lineno}
 
\journal{Renewable \& Sustainable Energy Reviews}

\begin{document}
\begin{frontmatter}




\title{Applications of blockchain and artificial intelligence technologies for enabling prosumers in smart grids: A review}

\author[label1]{Weiqi~Hua}
 \ead{huaw5@cardiff.ac.uk}
\author[label2]{Ying~Chen}
 \ead{cynthia.cy.chen@kpmg.com}
\author[label3]{Meysam~Qadrdan}
 \ead{qadrdanm@cardiff.ac.uk}
\author[label4]{Jing~Jiang}
 \ead{jing.jiang@northumbria.ac.uk}
\author{Hongjian~Sun\corref{cor1}\fnref{label5}}
 \ead{hongjian.sun@durham.ac.uk}
\author[label3]{Jianzhong~Wu}
 \ead{wuj5@cardiff.ac.uk}
 
\address[label1]{Department of Engineering Science, University of Oxford, Oxford OX1 3QG,UK}
\address[label2]{KPMG Huazhen LLP Shanghai Branch, 200010, Shanghai, China}
\address[label3]{School of Engineering, Cardiﬀ University, Cardiﬀ CF24 3AA, UK}
\address[label4]{Department of Mathematics, Physics and Electrical Engineering, University of Northumbria, Newcastle NE1 8ST, UK }
\address[label5]{Department of Engineering, Durham University, DH1 3LE, Durham, UK}

\cortext[cor1]{Corresponding author.}



\begin{abstract}
Governments' net zero emission target aims at increasing the share of renewable energy sources as well as influencing the behaviours of consumers to support the cost-effective balancing of energy supply and demand. These will be achieved by the advanced information and control infrastructures of smart grids which allow the interoperability among various stakeholders. Under this circumstance, increasing number of consumers produce, store, and consume energy, giving them a new role of prosumers. The integration of prosumers and accommodation of incurred bidirectional flows of energy and information rely on two key factors: flexible structures of energy markets and intelligent operations of power systems. The blockchain and artificial intelligence (AI) are innovative technologies to fulfil these two factors, by which the blockchain provides decentralised trading platforms for energy markets and the AI supports the optimal operational control of power systems. This paper attempts to address how to incorporate the blockchain and AI in the smart grids for facilitating prosumers to participate in energy markets.  To achieve this objective, first, this paper  reviews how policy designs price carbon emissions caused by the  fossil-fuel based generation so as to facilitate the integration of prosumers with renewable energy sources. Second, the potential structures of energy markets with the support of the blockchain technologies are discussed. Last, how to apply the AI for enhancing the state monitoring and decision making during the operations of power systems  is introduced.
\end{abstract}


\end{frontmatter}



\section{Introduction}
The power systems represent around 40 \% of global carbon emissions from the combustion of fossil fuels \cite{jiang2019structural}. In efforts to meet the targets of net zero power systems, policy
makers formulate measures for facilitating the integration of renewable energy sources (RESs) and encouraging changes in energy consumption behaviours. The smart grids refer to an intelligent power network which cost-effectively integrates information and control infrastructures to allow more reliable and efficient operations of power systems \cite{farhangi2009path}.  From the perspective of information system infrastructure, the smart grids enable bidirectional communications between stakeholders in power systems such as the system operator, generators, and consumers, which facilitates the optimal operation of generators and the active engagement of consumers \cite{yan2012survey}. From the control perspective, the interoperability of the smart grids enables the optimal coordination of various entities such as generation units or loads, to cooperatively achieve the overall benefits of power systems  \cite{basso2012update}.

The regulatory supports and advances of smart grids enable consumers to actively produce, consume, and store energy through using distributed RESs, storage devices, and advanced metering infrastructures. The energy markets are transitioning to recognise and promote the emerging role of prosumers \cite{parag2016electricity}. The prosumers are small or medium sized energy users \cite{shandurkova2012prosumer}, e.g., residents, businesses, and industries, who also generate energy on-site, and strategically exchange energy with the utility grid or other prosumers to meet their own demand or make profits from the energy arbitrage. The emerging role of prosumers is expected to promote the demand side management (DSM) and therefore reduce the dependency on the fossil-fuel based generation with the long-distance transmission. Nevertheless, the involvement of prosumers also brings the following challenges from the perspectives of energy markets and power systems:

$\bullet$ The structures of current energy markets are not suitable to accommodate the role of prosumers, since energy pricing schemes and balancing mechanisms are independent of the behaviours of energy exchange among prosumers \cite{paudel2018peer, zhou2020framework}.

$\bullet$ The information infrastructures of the current power systems cannot handle the increasing information flows caused by the decision making and transactions of large amounts of distributed prosumers \cite{ahl2019review}.

$\bullet$ Given limited budgets for small or medium sized prosumers' control systems, it is hard for them to exploit historical data for optimally scheduling the generation and consumption according to their specific energy patterns \cite{dupont2012linear}.

$\bullet$ It is challenging to accurately predict prosumption behaviours given uncertainties caused by the intermittency of distributed RESs and flexible demand \cite{MAHMUD2020109840}.

\begin{figure}[H]
	\centering	
	\includegraphics [width=6 in]{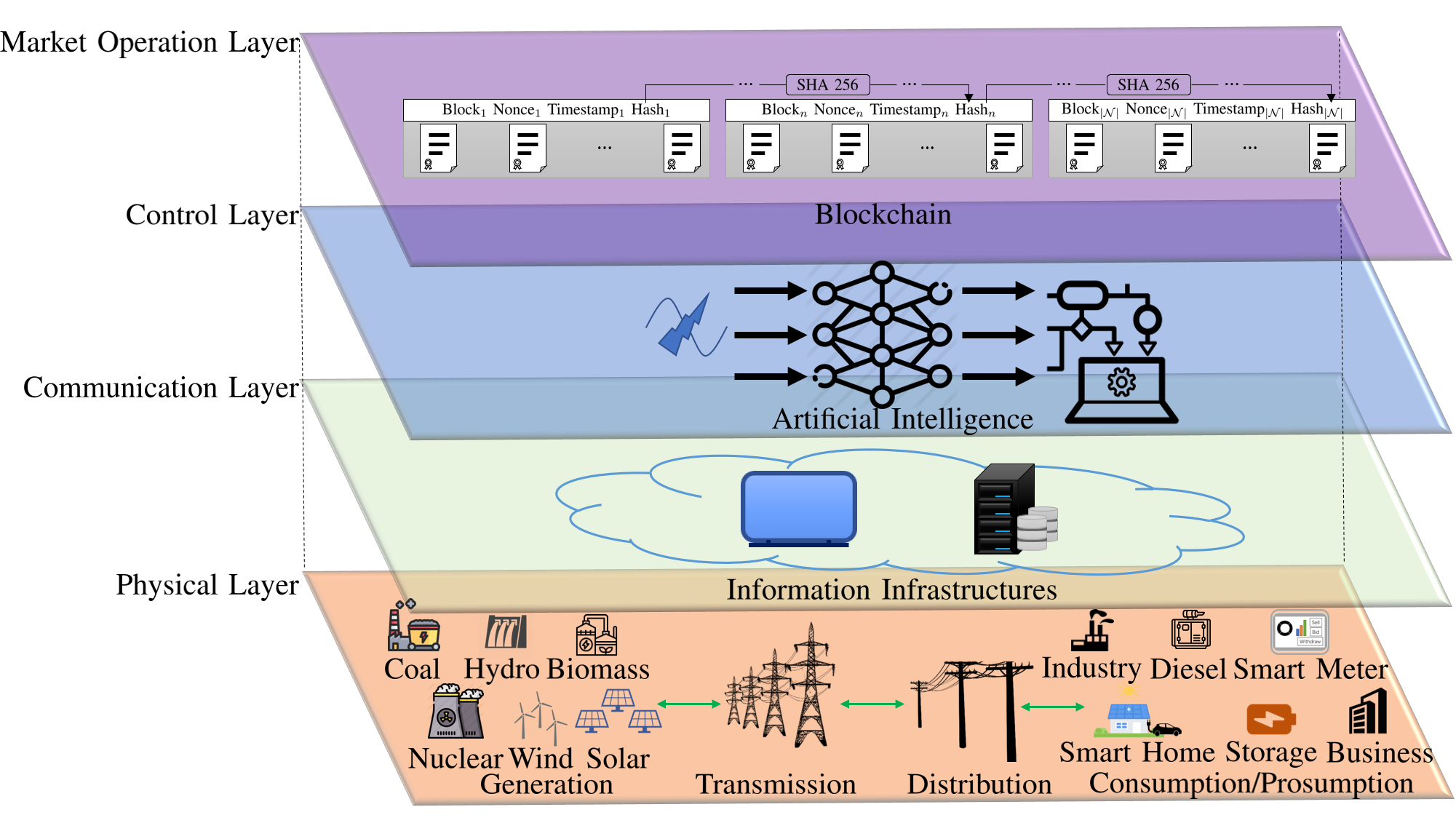}			
	\caption{Conceptual graph of incorporating the blockchain and artificial intelligence in smart grids.} \label{C2FIG1}	
\end{figure}

Flexible structures of energy markets and intelligent operations of power systems are two crucial factors for addressing these challenges.  These two factors can be fulfilled by recent scientific innovations on the blockchain and artificial intelligence (AI). From the perspective of the energy market, the blockchain provides the trading platform and technical supports for decentralised energy markets which are open and accessible to individual prosumers with the enhanced automation, security, and privacy \cite{thomas2019general}. From the operational perspective, the AI supports control systems to strategically make decisions for optimising system operations and achieving certain goals \cite{cheng2019new}, such as saving electricity bills, improving generating profits, mitigating carbon emissions, and predicting system uncertainties. The decisions are yielded by intelligent controlling approaches, such as the optimisation, game theory, and machine learning, which can take advantage of historical data from power systems. The conceptual graph of incorporating the blockchain and AI in smart grids with the prosumers' integration is presented in Fig. \ref{C2FIG1}.

The review presented in this paper is inspired by the issue of how the applications of blockchain and AI in smart grids could enable the integration of prosumers to decarbonise power systems. The rest of this paper is organised as follows: 
From the regulatory aspect, Section \ref{S2} introduces how to trace carbon emissions and then impose the carbon cost on fossil-fuel based generation while promote the engagement of prosumers with the distributed RESs.  From the aspect of market structures,  Section \ref{S3} identifies the potential structures of energy markets when integrating the role of prosumers, and reviews the research on blockchain implemented in decentralised energy trading. From the operational aspect, Section \ref{S4} reviews the research on how the AI supports the control and decision making of stakeholders in power systems. Section \ref{S5} draws the conclusion of this paper.

\section{Carbon emissions tracing and pricing}\label{S2}
This section reviews approaches for carbon emissions tracing in power systems as a foundation to inform the policy design, and then discusses how international regulations and existing research price carbon emissions from the fossil-fuel based energy generation.
\subsection{Carbon emissions tracing in power systems}
 In the context of this review paper, the term of \emph{carbon emissions} refers to the carbon dioxide equivalent which is a metric to measure the emissions from various greenhouse sources by converting the amounts of these sources to the equivalent  amount of the carbon dioxide according to their global warming potentials \cite{babacan2020assessing}.  The carbon emissions from power systems can be accounted by the carbon intensities or carbon emission flows (CEFs). The carbon intensities quantify the amount of carbon emissions per unit of energy generation by evaluating the carbon content of fossil fuels and generating efficiency \cite{zhao2012carbon}. The CEFs trace carbon emissions from generation when the power is transmitted and consumed \cite{kang2012carbon}, which is crucial for fairly allocating the responsibilities of carbon reduction and encouraging the demand side engagement. 
\subsubsection{Carbon intensities}
Evaluation of carbon intensities has been focused by a majority of research \cite{report, thomson2017marginal, siler2012marginal, hawkes2010estimating}. The research in \cite{thomson2017marginal} investigated the relationship between the dynamic carbon  intensities and the levels of the part-load operation of fossil-fuel based power stations using  historical data from power systems. The marginal generators, e.g., coal and gas, are generators which respond to the changes of RESs outputs by operating at the part-load state \cite{thomson2017marginal}. The part-load operation of the marginal generators would reduce efficiencies, which consumes more fossil fuels and raises the carbon intensities.  To further investigate the impacts of this part-load operation on the carbon emissions, researchers analysed the historical data of all generation sources \cite{report}, marginal generators  \cite{siler2012marginal}, and demand \cite{li2020can}, through which three corresponding types of emission factors were defined: the average emission factor, marginal displacement factor, and marginal emission factor. The average emission factor quantifies the part-load impacts on the annual average carbon emissions from all power generation sources; The marginal displacement factor quantifies the part-load impacts on the carbon emissions from generators operating at the margin; The marginal emission factor quantifies  the part-load impacts on the carbon emissions from the marginal changes of the power demand.  Moreover, to audit the carbon emissions caused by the RESs, the long-term average carbon intensities evaluated by the life-cycle carbon analysis \cite{kneifel2010life} are used. 
\subsubsection{Carbon emission flows}
To facilitate the demand side engagement for decarbonising power systems, the responsibilities of carbon emissions from  generators can be shared by consumers and prosumers, since the consumption and power import from the utility grid are the primary driver to the fossil-fuel based generation. Sharing the responsibilities of carbon emissions requires the information of how much carbon emissions are produced by generators when transmitting and consuming per unit of energy. This information can be obtained by analysing the topological structures and power flows of power networks using the CEFs. The CEFs are virtual network flows concurrent with the power flows to analyse the responsibilities of carbon emissions for every component of power networks including transmission lines and loads  \cite{kang2012carbon}. The approach of CEFs has been focused in the literature. The concept of CEFs was initially created from international trades to audit  carbon responsibilities among countries. St{\aa}hls \emph{et al.} \cite{staahls2011impacts} analysed the international carbon flows from a consumption-based perspective and identified the portion of carbon emissions from industrial exports. Further research implemented this concept into power systems to identify the carbon emissions incurred by consumption behaviours. In \cite{li2013carbon}, an approach was developed for analysing the CEFs to determine the indirect carbon emissions caused by consumption behaviours, by which the regional variation of carbon emissions was identified. Kang \emph{et al.} \cite{7021901} quantified the carbon emissions from the power delivery process by analysing the operational characteristics and topological structures of power networks.

\subsection{Policy design for pricing carbon emissions}
The carbon pricing is a market based policy to address carbon emissions caused by the combustion of fossil fuels \cite{fowlie2016market}. This policy enforces the pollutant emitters to compensate the environmental damage in a monetary manner, which increases the costs of using fossil fuels and subsequently encourages the engagement of prosumers with their distributed RESs. Two primary forms of carbon pricing are the carbon tax and emissions trading scheme. By the end of 2019, the policy of carbon pricing has been implemented in 46 countries, of which 25 countries adopt the carbon tax and 21 countries adopt the emissions trading scheme \cite{ramstein2019state}.

\subsubsection{Carbon tax}
The carbon tax levies a fixed rate on carbon content of fossil fuels \cite{metcalf2009design}. This fixed rate is determined by the social cost of carbon which quantifies the marginal damage costs of carbon emissions to the society \cite{nordhaus2017revisiting}.  
\subsubsection{Emissions trading scheme}
The emissions trading scheme, also known as the cap-and-trade scheme, is an alternative form to the carbon tax. Under the emissions trading scheme, the policy makers allocate a certain amount of carbon allowances for a given time period \cite{driga2019climate}. Carbon producers are obliged to have an enough amount of carbon allowances covering the amount of their carbon emissions. The surplus or deficiency of carbon allowances can be traded among these carbon producers \cite{stavins2003experience}. 

Nonetheless, an inappropriate carbon price determined by the emissions trading scheme would inefficiently deliver the targets of carbon reduction. If the carbon price lies below the social cost of carbon or the rate at which the targets of carbon reduction can be achieved, it would insufficiently stimulate the mitigation of carbon emissions; If the carbon price in one region is higher than the carbon price in another region, the market competitiveness  of carbon producers in the high-price region would be harmed. The carbon producers are prone to discharging carbon emissions in the low-price region, while the total amount of carbon emissions remains unchanged, which is defined as the carbon leakage issue \cite{babiker2005climate}. 

To overcome the issue of an inappropriate carbon price, the carbon price floor and ceiling are implemented in current international carbon markets by setting additional limits to carbon prices \cite{doda2016price}. For the case of the U.K. carbon market, since the carbon price of the E.U. emissions trading scheme is lower than the social cost of carbon in the U.K., the carbon price had failed to incentivise the U.K. coal-to-gas transition before 2013 \cite{carbonpricefloor}. Afterwards, the U.K. has formulated the carbon price floor as the lower bound of the carbon prices from the E.U. emissions trading scheme. The U.S. has set a similar price floor and facilitated the carbon auctions since 2009 \cite{goulder2013carbon}. In New Zealand, a carbon price ceiling has been enacted by a fixed price option to prevent high carbon prices and protect the market competitiveness of domestic carbon producers \cite{newbery2019political}.

\begin{figure}[H]
	\centering
	\includegraphics [width=4.5in]{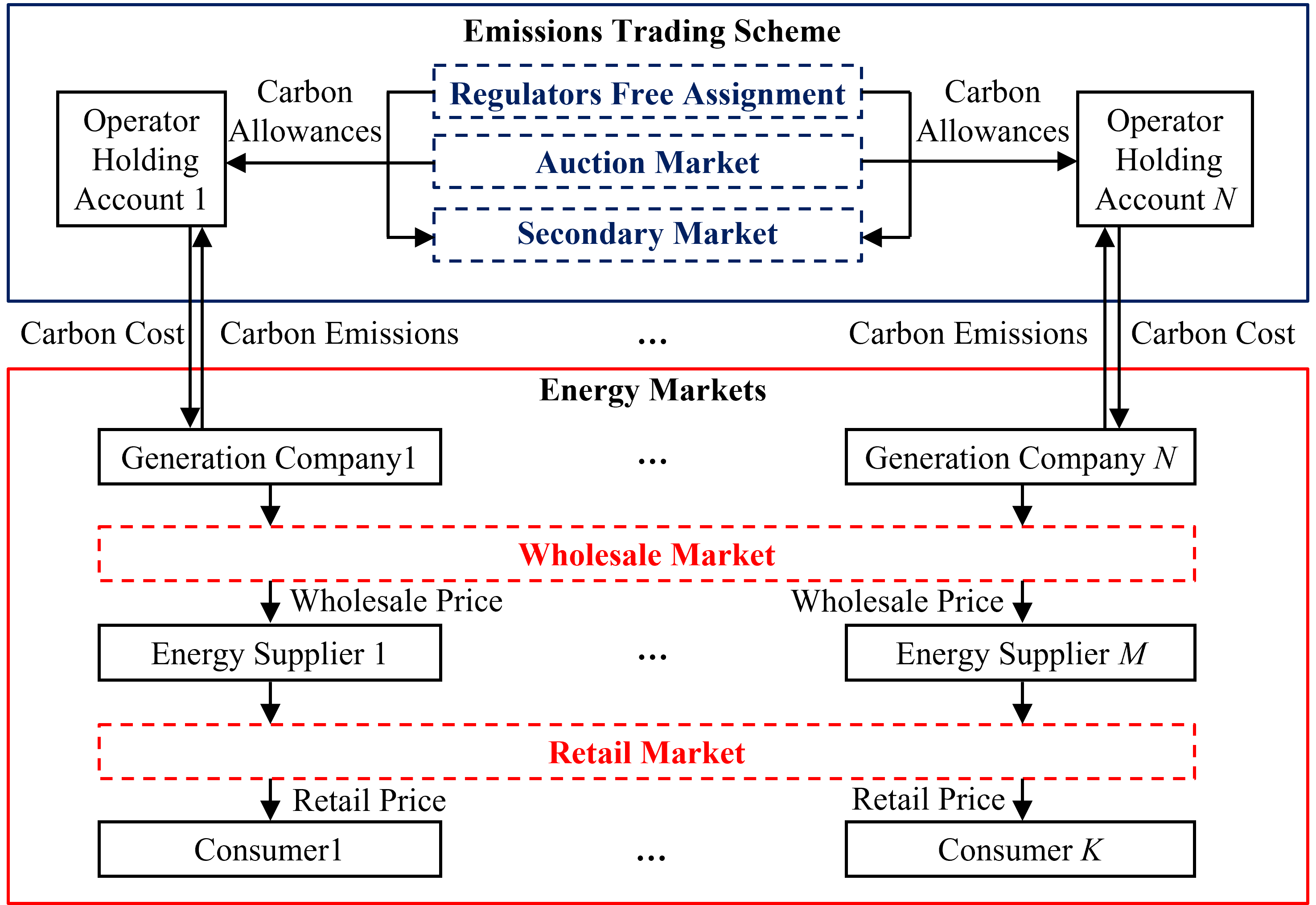}
	\caption{Relationship between the emissions trading scheme and energy markets.} \label{FIG0}	
\end{figure}

The relationship between the emissions trading scheme and energy markets is presented in Fig. \ref{FIG0}. The emissions trading scheme is linked to the energy markets through the carbon emissions and incurred carbon cost of generation companies.  As a key stakeholder in energy markets, generation companies need to register an operator holding account in order to participate in the carbon markets \cite{ETSch}. With this operator holding account, generation companies can \emph{1)} receive free carbon allowances from regulators if they are eligible, \emph{2)} bid carbon allowances from the auction market held by regulators, \emph{3)} buy/sell carbon allowances from/to other pollutant emitters with surplus/deficiency of carbon allowances at the secondary market, and \emph{4)} prove to regulators that their carbon allowances cover their reportable emissions.

Purchasing carbon allowances, i.e., carbon cost, is a part of operational costs of generation companies, and the carbon cost varies depending on the carbon intensities of different generation technologies as 
\begin{equation}\label{eq1}
	c^{\textup{carbon}}_{n}= \pi ^{\textup{carbon}}\cdot \sum_{g\in G}\left [i_{g,n}\cdot  \sum_{t\in T} \left ( p_{g,n,t}\cdot \Delta  t\right )\right ],
\end{equation}
where $c^{\textup{carbon}}_{n}$ is the carbon cost of the generation company $n$, $\pi ^{\textup{carbon}}$ is the carbon price determined at the auction or secondary markets with the unit of GBP/ton, $i_{g,n}$ is the carbon intensity of the power plant $g$ belonging to the generation company $n$ with the unit of ton/MWh, $G$ is the index set of all power plants  belonging to the generation company $n$,  $p_{g,n,t}$ is the power output of the power plant $g$ belonging to the generation company $n$ at the time step $t$ with the unit of MW, $T$ is the index set of time steps, and $\Delta  t$ is the time interval.

The life-cycle carbon intensities of different generation technologies are compared  in Table \ref{TABLE00}. It can be seen from the table that the carbon intensities of fossil-fuel based generation	technologies are higher than those of RESs, and therefore the generation companies with fossil-fuel based generation need to afford more carbon cost according to Eq. (\ref{eq1}). In the energy market, the carbon cost is further passed from generation companies to consumers through wholesale energy markets and retail energy markets (see Fig. \ref{FIG0}), in the form of increased electricity bills.

\begin{table}[H]\scriptsize
	\caption{Comparison  of life-cycle carbon intensities of different generation technologies.}
	\centering
	\label{TABLE00}
	\begin{tabular}{lllllllll}
		\hline	
	Carbon Intensity (ton/MWh)	&Coal&Gas &Biomass & Solar PV&Hydro&Wind Onshore&Wind Offshore&Nuclear  \\	\hline	
	IPCC \cite{annex2014metrics}   &0.820 &0.490 &0.230&0.048&0.024&0.011&0.012&0.012\\		
	UNECE  \cite{unecereport}    &1.000 &0.430 & -&0.037&0.011&0.012&0.013& 0.005 \\			
	
		\hline			
	\end{tabular}
\end{table}

\begin{figure}[H]
	\centering
	\includegraphics [width=4.5in]{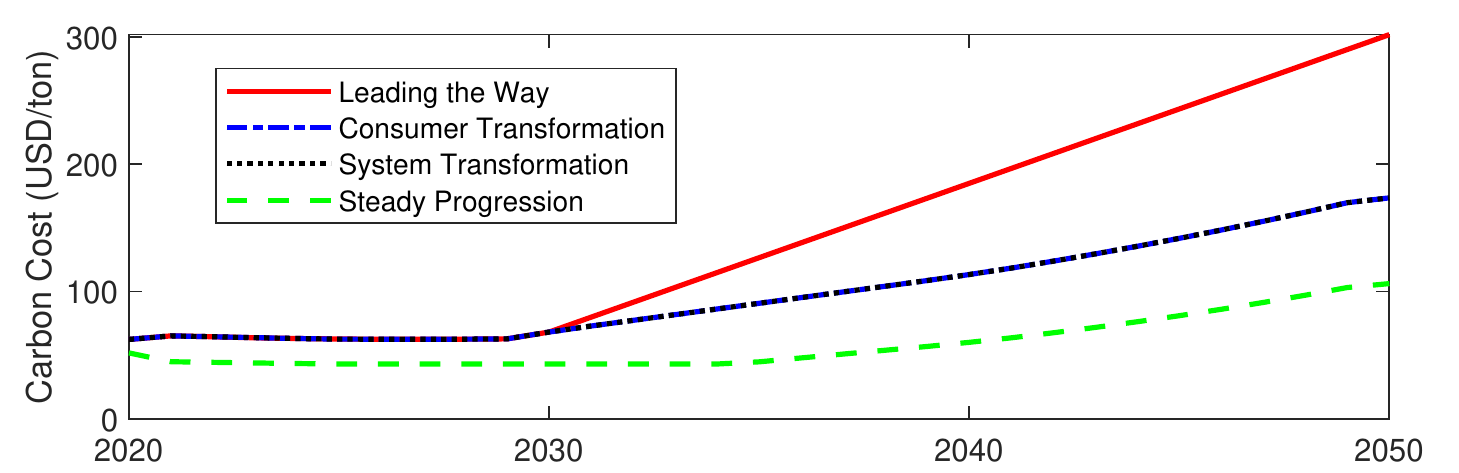}
	\caption{The projection of the future UK carbon cost under the UK emissions trading scheme, which includes  four scenarios of societal changes in achieving the 2050 net zero target identified by the nationalgridESO future energy scenarios 2020 \cite{carboncossst}.} \label{carboncost}	
\end{figure}

The projection of the future UK carbon cost under the UK emissions trading scheme is presented in Fig. \ref{carboncost}, which includes  four scenarios of societal changes in achieving the 2050 net zero target identified by the nationalgridESO future energy scenarios 2020 \cite{carboncossst}. The scenario of the \emph{steady progression} indicates the slowest path to the decarbonisation compared to other scenarios, under which the energy supply would heavily rely on the natural gas and domestic heating, while there will be slight increase of the home insulation and uptake of electric vehicles. Due to the relatively low carbon cost, this scenario would fail to achieve the net zero target by 2050. The scenario of the \emph{system transformation} indicates a significant transformation on the supply side while slight changes on the consumers, whereas the scenario of the \emph{consumer transformation} indicates a significant level of consumers' engagement through the high penetration of low carbon heating sources, electric vehicles, smart energy managements, and storage devices. Both scenarios of the system transformation and consumer transformation have the same trend of the carbon cost and will achieve the net zero target by 2050. The scenario of the \emph{leading the way} indicates the most progressive path to the net zero target, by which different areas will achieve decarbonisation in their earliest dates with the highest consumers' engagement,  improvement of energy efficiency, and investment in low carbon technologies, and therefore this scenario yields the highest carbon cost.

Given the link between the emissions trading scheme and energy markets, separately designing the carbon and energy markets would be inefficient. Recent research has highlighted the importance for coupling the energy and carbon markets. Zhang \emph{et al.} \cite{zhang2015unit} integrated the emissions trading scheme into the unit commitment problem of electricity generators, enabling generation companies to trade energy and carbon allowances simultaneously. Huang \emph{et al.} \cite{huang2014experimental} analysed how the emission price, emission quality, and time related factors drive energy generation companies to participate in the emissions trading scheme. Researchers in \cite{skarvelis2013implementing} implemented the EU emissions trading scheme to control the emissions from a group of micro-generators.
	
Since increasing numbers of prosumers participate in the peer-to-peer energy trading and take the responsibilities of carbon reduction, designing decentralised emissions trading scheme has drawn attentions from recent research. Fawcett \cite{fawcett2010personal} reviewed the personal carbon trading undertaken by the UK government in 2008 which assigns individuals a tradable carbon allowance to cover emissions from personal energy consumption, and highlighted  its importance for the individual and social change in terms of carbon reduction.  A peer-to-peer trading framework was developed in  \cite{HUA2020115539} enabling prosumers to trade energy and carbon allowances together, in which a decentralised carbon incentive was formulated targeting on specific energy behaviours of prosumers to achieve the local energy balance and carbon reduction. Yan \emph{et al.} \cite{9525436} proposed a blockchain based transacting energy and carbon allowance between microgrids and the  distribution system operator.

\subsubsection{Comparison remark}
	As two well-established policy instruments, the carbon tax and emissions trading scheme have following aspects in common:

$\bullet$ Both the carbon tax and emissions trading scheme impose a price on carbon emissions for facilitating carbon producers to internalise the cost of environmental damages. 

$\bullet$ Instead of the command-and-control based policy that sets out and enforces specific actions for carbon reduction, the market based policy can flexibly incentivise carbon producers for strategically responding to the carbon prices.

$\bullet$ Market based policy can generate public revenue through charging the carbon tax or selling the carbon allowances. This revenue can be further redistributed for investing in low carbon technologies such as RESs and the carbon capture and storage, so as to achieve the carbon revenue neutrality \cite{gillis2019carbon}.

	The differences between the carbon tax and emissions trading scheme are as follows:

	$\bullet$ The carbon tax gives a certainty to the price of carbon emissions through a fixed tax rate, whereas the emissions trading scheme gives a certainty to the quantity of carbon emissions through the fixed total carbon allowances \cite{cvt}.
	
	$\bullet$ The carbon tax is easier to be implemented since it is based on the established tax systems. By contrast, the emissions trading scheme is more flexible since it can embed financial innovations, e.g., peer-to-peer trading and options.

\subsection{Remark of research challenges}
Although the carbon pricing has been implemented as practical regulations and investigated in the literature, there are still opportunities to incorporate dynamic and decentralised policy measures targeting on high-carbon generators and consumers. This is because the long-term policy for overall power systems cannot specifically target on real-time power profiles and incurred carbon emissions. 

Furthermore, with the increasing engagement of prosumers into local energy markets, tracing carbon emissions caused by individual energy patterns presents a challenge. These energy patterns are reflected by how a prosumer responds to pricing incentives through determining  its on-site generation, consumption, and energy exchange. Tracing prosumer-centric carbon emissions is particularly important when assigning personal carbon allowances to individual prosumers.

\section{Energy markets transition with prosumers' integration}\label{S3}
This section identifies the  potential structures of decentralised energy markets with the integration of the role of prosumers, and then reviews the research and innovations on how to exploit the blockchain technologies including smart contracts for facilitating the decentralised energy trading.

\subsection{Potential structures of decentralised energy markets} 
\begin{figure}[H]
	\centering	
	\includegraphics [width=6 in]{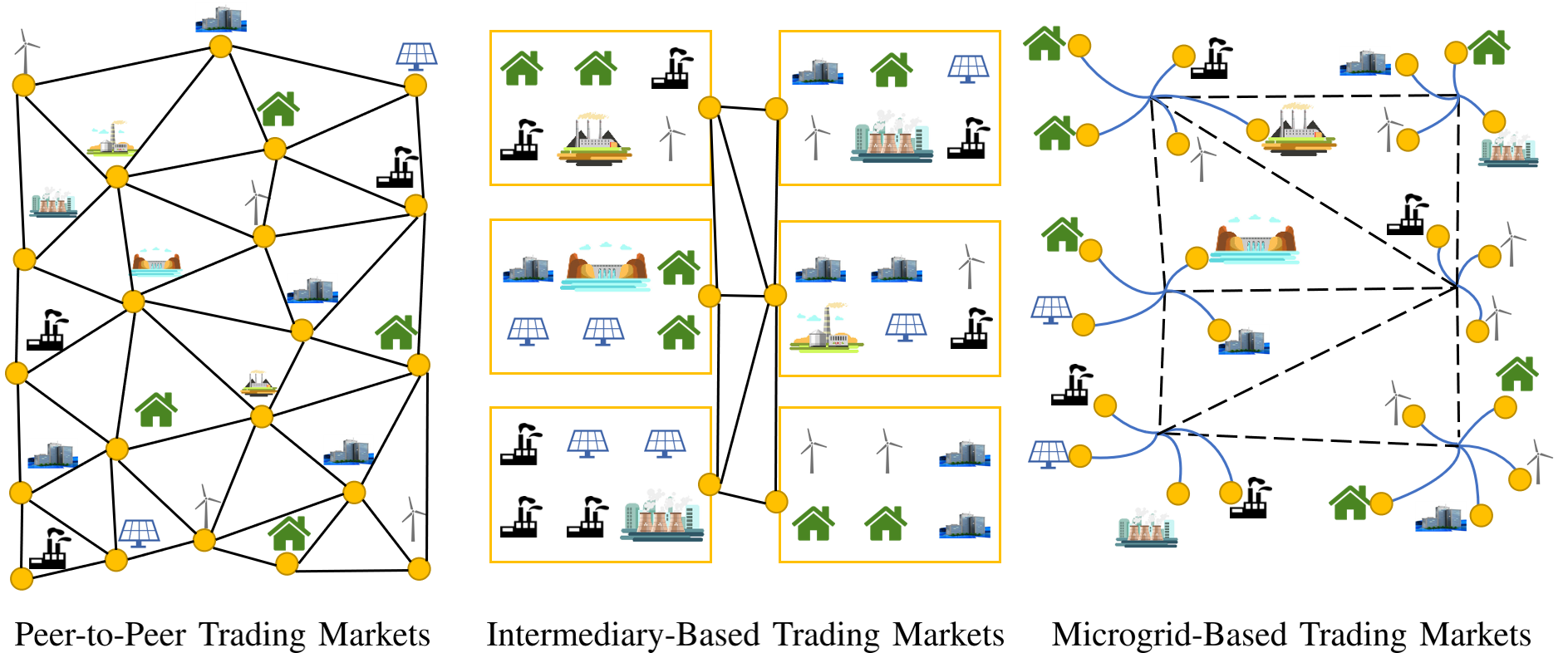}			
	\caption{Schematic illustration for structures of decentralised energy markets with the integration of the role of prosumers. Dots indicate the control units. Lines indicate the information flows exchanged among these units. Under the peer-to-peer trading markets, prosumers interconnect with each other to trade energy and other services; Under the intermediary-based trading markets, an ensemble of prosumers is organised by an intermediary to pool generation sources, flexible demand, and storage capacities for collective control; Under the microgrid-based trading markets, prosumers connect to microgrids and microgrids either connect to the utility grid or operate in an islanded mode as indicated by dashed lines.} \label{C2FIG2}	
\end{figure}

\begin{table}[H]\scriptsize
	\caption{Comparison for structures of decentralised energy markets with the integration of the role of prosumers.}
	\centering
	\label{TABLE0}
	\begin{tabular}{llll}
		\hline	
		&\!\!\!\! Peer-to-peer trading markets&\!\!\!\! Intermediary-based trading markets&\!\!\!\! Microgrid-based trading markets \\	\hline	
		Structure    &\!\!\!\! Prosumer-centric &\!\!\!\! Community-centric&\!\!\!\! Prosumer to microgrid to utility grid (or islanded) \\		
		Control unit      &\!\!\!\! Prosumers&\!\!\!\! Intermediary&\!\!\!\! Prosumers  \\			
		\multirow{2}*{Objective}      &\!\!\!\! \multirow{2}*{Individual prosumers' benefits}&\!\!\!\! \multirow{2}*{Community's benefits}&\!\!\!\! Profits for exporting (when connected to utility grid) \\        
		& \!\!\!\! &\!\!\!\! &\!\!\!\! Community's benefits (when islanded) \\        				
		\multirow{2}*{Pricing scheme}   &\!\!\!\! \multirow{2}*{Prosumers' bidding/selling prices}&\!\!\!\! \multirow{2}*{Intermediary's bidding/selling prices}       &\!\!\!\!  Retail prices (when connected to utility grid)  \\    
		&\!\!\!\! &\!\!\!\!     &\!\!\!\! Microgrid's bidding/selling prices  (when islanded)   \\   
		
		Implementation       &\!\!\!\! RWE \cite{RWE},  Power Ledger \cite{pl} &\!\!\!\! Stem \cite{Stem},  Energy and Meteo Systems  \cite{emsys}           &\!\!\!\! Asea Brown Boveri Ltd \cite{ABB}, LO3 Energy \cite{lo3} \\
		Advantage   &\!\!\!\! Fully decentralised &\!\!\!\! Coordinated within the community&\!\!\!\! Decentralised and coordinated within the microgrid \\

	Disadvantage    &\!\!\!\! High burdens of information and control &\!\!\!\! Centralised by the intermediary&\!\!\!\! Difficulty of aligning individual prosumers' profits\\	
				  &\!\!\!\!  &\!\!\!\! &\!\!\!\! with microgrid benefits \\
		Structure&\!\!\!\! Least structured framework&\!\!\!\! Moderate structured framework&\!\!\!\! Most structured framework\\
				difference&\!\!\!\! &\!\!\!\! &\!\!\!\! \\

		Common&\multicolumn{3}{l}{\!\!\!\! Flexible structures of decentralised energy markets which accommodate increasing burdens of information and control incurred by}\\
		& \multicolumn{3}{l}{\!\!\!\! the engagement of prosumers}\\
	
		\hline			
	\end{tabular}
\end{table}
A transition of energy markets towards decentralised generation and consumption is crucial for the integration of the emerging role of prosumers. The potential structures of such markets have been well investigated \cite{parag2016electricity} and three primary structures are identified: \emph{1)} peer-to-peer trading markets, \emph{2)} intermediary-based trading markets, and \emph{3)} microgrid-based trading markets.  These three structures of energy markets  are based on the information and control infrastructures of smart grids, and categorised by the functions of control units and associated manners of the information exchange. A schematic illustration of these three structures is presented in Fig. \ref{C2FIG2}, where each dot represents a control unit and each interconnected line represents an information flow. The comparison of these three structures is presented in Table \ref{TABLE0} with details introduced as follows.

\subsubsection{Peer-to-peer trading markets}
The peer-to-peer trading markets are structured as a completely decentralised framework \cite{hamari2016sharing}, under which the energy and services, e.g., DSM, storage capacities, and carbon credits, can be directly traded among prosumers. In comparison to the other two market structures, the peer-to-peer trading markets are the least structured framework. Instead of using central authorities, e.g., aggregator, as a control unit, each individual prosumer becomes an independent unit to perform control functions and exchange information with each other \cite{8279516}.  The behaviours of  prosumers are directly incentivised by their individual bidding/selling prices. The role of the distribution system operator remains as managing the trading platform and providing the power distribution function \cite{PARK2019324}. Hence, this framework allows individual prosumers to directly participate in energy markets while increases the burdens of control and information flows.

As practical cases, the RWE \cite{RWE} has developed the peer-to-peer trading platforms integrating the functions of the decentralised generation control, grid management, communication, automation, and security. The Power Ledger \cite{pl} provides a software based peer-to-peer energy trading for 11,000 participants from residential and commercial consumers in Australia.
\subsubsection{Intermediary-based trading markets}
The intermediary-based trading markets are more structured than the peer-to-peer trading markets, under which an ensemble of prosumers is organised as a community such as smart buildings and virtual power plants. 
Each community is managed by an intermediary, e.g., aggregators or retailers, as a control unit to perform control functions and exchange information with each other. All generation sources, loads, and storage capacities within a community are pooled to collectively coordinate resources for local benefits. The intermediary can earn  bonus from regulators or utilities by providing prosumers with services, e.g., the improvement of residential energy efficiency, DSM, and setup of RESs. 

An example of the intermediary is the Stem \cite{Stem} which has designed a platform to provide the storage service and DSM for consumers in California through the real-time optimisation and automated control. The company of Energy and Meteo Systems  \cite{emsys} in Germany has established a virtual power plant via a digital control centre with the services of the real-time data management, remote control of wind and solar generation, energy scheduling, DSM, and balancing management. 
\subsubsection{Microgrid-based trading markets}
The microgrid-based trading markets are the most structured framework, under which prosumers connect to microgrids and microgrids can either connect to utility grids or operate in an islanded mode \cite{kahrobaeian2012interactive}. Analogous to the peer-to-peer trading markets, each individual prosumer is an independent control unit connecting to the microgrid without the intermediary. When a microgrid connects to the utility grid, prosumers can sell surplus generation to the utilities \cite{parag2016electricity}. In this case, prosumers would be incentivised to generate more energy for earning profits.  When a microgrid operates in an islanded mode, the surplus generation can be stored within the microgrid or used for load shifting services \cite{di2014near}. In this case, prosumers would be incentivised to strategically schedule their generation and consumption for the local energy balance. 

As practical cases, the Asea Brown Boveri Ltd \cite{ABB} provides microgrid solutions for customers to ensure the reliable, stable, and affordable power supply.  The LO3 Energy \cite{lo3} has developed the Brooklyn microgrid integrating 130 buildings to facilitate the DSM and improve communication infrastructures. 
\subsection{Blockchain supporting decentralised energy trading}
This subsection introduces the concepts, advantages and limitations of blockchain technologies. The smart contracts which are the second generation of the blockchain technologies and the most potential application in the decentralised energy trading \cite{zhou2020state} are specifically focused. The research and implementations on applying the blockchain technologies in the decentralised energy trading are reviewed and subsequently compared to the conventional centralised trading.
\subsubsection{Blockchain Technologies}
Blockchain technologies  \cite{mengelkamp2018blockchain}, as one of the distributed ledger technologies, have the potential to establish a platform for the decentralised energy trading. The blockchain can prevent the replay attack and double spending attack \cite{karame2012double} in energy markets, i.e., the same energy is sold twice or the same digital currency is spent twice, through accounting the ownership of these assets. The decentralised feature of the blockchain  enables a ledger to be held and verified by all energy market participants \cite{karame2012double}. Hence, the trading platform is open and accessible for all prosumers, system operators, and market operators. The disintermediating feature of the blockchain transits the role of energy suppliers or aggregators to a neutral facilitator for encouraging prosumers' participation \cite{fairfield2014smart}.  The encryption of the blockchain protects prosumers' private information such as addresses, transactions, and power profiles. The computational difficulty of block mining and collective validation for reaching a consensus guarantee the security of trading networks \cite{al2003certificateless}.	

Nonetheless, due to the technical limitations of current blockchain technologies and the conflicts with physical assets of power systems, the application of blockchain technologies also brings challenges. First, theoretically the blockchain networks allow the prosumers at anywhere to trade energy with each other. However, this would violate the physical restriction of power systems and cause higher power losses over the long-distance transmission. How to ensure prosumers to trade energy within their distribution networks presents a challenge. Second, the throughput, i.e., transactions per second, of the blockchain is lower than the existing trading technologies, whereas the latency, i.e., time per verified transaction, of the blockchain is higher than the existing trading technologies. For instance, the throughputs of the Ethereum, Bitcoin, and Visa are 15 \cite{8342866},7 \cite{8855462}, and 2000 \cite{mcconaghy2014blockchain} transactions per second, respectively, whereas the latencies of them are 3 \cite{spain2020impact}, 10 \cite{shah2014bayesian}, and 0.05 \cite{wust2018you} minutes, respectively.
\subsubsection{Smart Contracts}
Blockchain technologies have evolved from the first generation of the Bitcoin and cryptocurrency to the second generation of the Ethereum and smart contracts. In the field of the decentralised energy trading, the most potential application of the blockchain technologies is the smart contracts. The smart contracts, coined by Szabo in 1994 \cite{christidis2016blockchains}, enable executable programs to be performed in a manner of the self-enforcing settlement and setting out negotiation \cite{buterin2014next}. This supports the automatic control and interoperability of the smart grids, so as to reduce the burdens of handling information exchanges among prosumers. A general form of the smart contracts is `\emph{If an event A happens, the smart contracts pay the currency B, deposited by the buyer C, to the seller D}' \cite{levi2018introduction}. The replicable feature of this general form of the smart contracts ensures standardised trading procedures with reduced transactional costs, and also prevents unforeseen trading behaviours. On the context of the energy trading, the event could be the supply of energy or other services, e.g., the DSM, which is monitored by smart meters of prosumers. The pay function is executed in a self-enforcing manner. Hence, the trustworthiness of the energy trading is dependent on the trustworthiness of smart meters and programs to be executed on the smart contracts.  Nevertheless, the interactions between the smart contracts and smart meters or controllers require the design of new communication protocols and interfacing domains.
\subsubsection{Research and implementations}
The blockchain and smart contracts applied in the power systems control and decentralised energy trading are the subject of active research and practical implementations. Thomas \emph{et al.} \cite{thomas2019general} proposed a general form of smart contracts  for negotiation and controlling energy transfer process between separated distribution networks. In \cite{di2018technical}, the real-time power losses caused by energy trading in microgrids were accounted by the blockchain, by which the prosumers were considered as negotiators of energy trading and distributors were responsible for computing losses. Li \emph{et al.} \cite{li2019design} applied smart contracts into distributed hybrid energy systems to facilitate the energy exchange among the end-user. The DSM and uncertainties caused by the renewable generation were considered into the designed framework of the peer-to-peer energy trading. Mihaylov \emph{et al.} \cite{6861213} designed a paradigm for the energy trading with a virtual currency generated from the energy supply. Case studies of this research testified that the designed currency incentivised prosumers to achieve the demand response and energy balance. Saxena \emph{et al.} \cite{saxena2020blockchain} proposed a blockchain based transactive energy system to address the incentivising, contract auditability and enforcement of the voltage regulation service. The smart contracts were used by this research to enforce the validity of each transaction and automate the negotiation and bidding processes. In \cite{myung2018ethereum}, a transparent and safe energy trading algorithm executed on the Ethereum blockchain platform was presented.

To enhance the carbon pricing scheme, blockchain technologies have also been developed to trade the carbon allowances or allocate the incentives for decarbonisation. Khaqqi \emph{et al.} \cite{khaqqi2018incorporating} customised the trading of carbon allowances to industries using a reputation based blockchain network by which the reputation signified participants' performances and commitments for the carbon reduction.  Pan \emph{et al.} \cite{pan2019application} implemented the blockchain into the trading of carbon credits to reduce the entry threshold of carbon markets and improve the reliability of information exchange. Analogously, Richardson and Xu \cite{richardson2020carbon} proposed a blockchain based emissions trading scheme to ensure the transparency, tamper-resistance, and high liquidity. With respect to the application of smart contracts, a distributed carbon ledger system fitted with existing emissions trading schemes was designed in \cite{tang2019toward} to strengthen the corporate carbon accounting systems.

\subsubsection{Comparison between the centralised trading and blockchain based decentralised trading}
The difference between the conventional centralised trading and blockchain based decentralised trading in energy markets is summarised in Table \ref{TABLE22}, with detailed explanations as follows:

\begin{table}[H]\scriptsize
	\caption{Comparison between the conventional centralised trading and blockchain based decentralised trading in energy markets.}
	\centering
	\label{TABLE22}
	\begin{tabular}{lll}
		\hline	
		&Conventional centralised trading       &Blockchain based decentralised trading \\	\hline	
		Generators     &Large scale power plants  &Prosumers with distributed RESs \\		
		Pricing scheme       &Determined by wholesale or retail markets	&Prosumer-centric bidding/offering pricing  \\			
		Contract type        &Idiosyncratic contracts \cite{NGS}   &Standardised smart contracts   \\        
		Settlement enforcement \cite{buterin2014next}   &Legal restriction        &Self-enforcement       \\    
		Trustee \cite{buterin2014next}             &Third party                &Smart meters, smart contracts, and consensus \\
		
			Advantage  &Centralised coordination and negotiable contracts &Decentralisation, standardisation, and automation to prevent unforeseen \\
			 &&trading behaviours \\		
		
		Disadvantage   &Pricing and contract may not reflect individual &Pricing may not reflect supply-demand balance in overall energy markets,\\	
        & behaviours,  and dependence on a third party & and attacks to blockchain networks\\	
	 Objective difference&Designed for large-scale power plants&Designed for prosumers\\
		Common&\multicolumn{2}{l}{Incorporating pricing and regulatory mechanisms into energy markets to ensure the supply-demand balance and security of supply }\\			
		\hline			
	\end{tabular}
\end{table}

First, the primary generators in the conventional centralised trading are large-scale fossil-fuel based power plants connecting to  transmission networks, whereas the primary generators in the decentralised trading are prosumers with distributed RESs connecting to distribution networks. 

Second, the pricing scheme in the conventional energy trading reflects the supply-demand balance of overall energy markets. For instance, the wholesale electricity prices are determined by the uniform market clearing pricing or pay-as-bid pricing \cite{kahn2001uniform}. By contrast, the prosumer-centric pricing scheme in the decentralised energy trading can reflect individual supply-demand balance.

Third, the contracts for the conventional energy trading are idiosyncratic \cite{NGS}, which means that the contents of  contracts are negotiated between generators, suppliers, system operators, market operators, and policy makers.  By contrast, the smart contracts formulate standardised auction procedures for the decentralised energy trading, which is replicable for all prosumers, which can prevent unforeseen trading behaviours of prosumers

Fourth, the settlement of the centralised energy trading is enforced by legal restrictions, which means that if the energy or other services are not delivered, generators or suppliers would receive penalty afterwards. By contrast, the self-enforcing settlement of smart contracts enables the violation of contracts to be prevented beforehand by querying smart meters to ensure that the prosumers have enough capacities to supply.

Fifth, the trustworthiness of the conventional  energy trading relies on a third party, e.g., the auditing institutions or market operators, whereas in the blockchain based decentralised energy trading, the trustworthiness of prosumers relies on the consensus of blockchain networks and the interface between smart contracts and smart meters.
\subsection{Remark of research challenges}
Although these innovative structures of energy markets and the blockchain based decentralised energy trading can support the integration of prosumers, the transition of energy markets also raises a series of challenges as follows:

First, when prosumers feed their distributed generation into the utility grid, the issues on market operations, e.g., negative energy prices \cite{nicolosi2010wind}, and grid operations, e.g., the voltage spike \cite{he2018high}, power imbalance \cite{qiu2017optimal}, and harmonic distortion \cite{zhao2020decentralized}, would challenge the control infrastructures and protocols of current power systems.

Second, for these decentralised energy markets without central authorities, how to maintain the overall benefits of power systems, e.g., the resilience and carbon mitigation, presents a challenge. This requires sophisticated rulesets, incentive measures, and pricing schemes to align individual prosumers' behaviours with systems' benefits. 

\begin{figure}[H]
	\centering
	\includegraphics [width=4.5in]{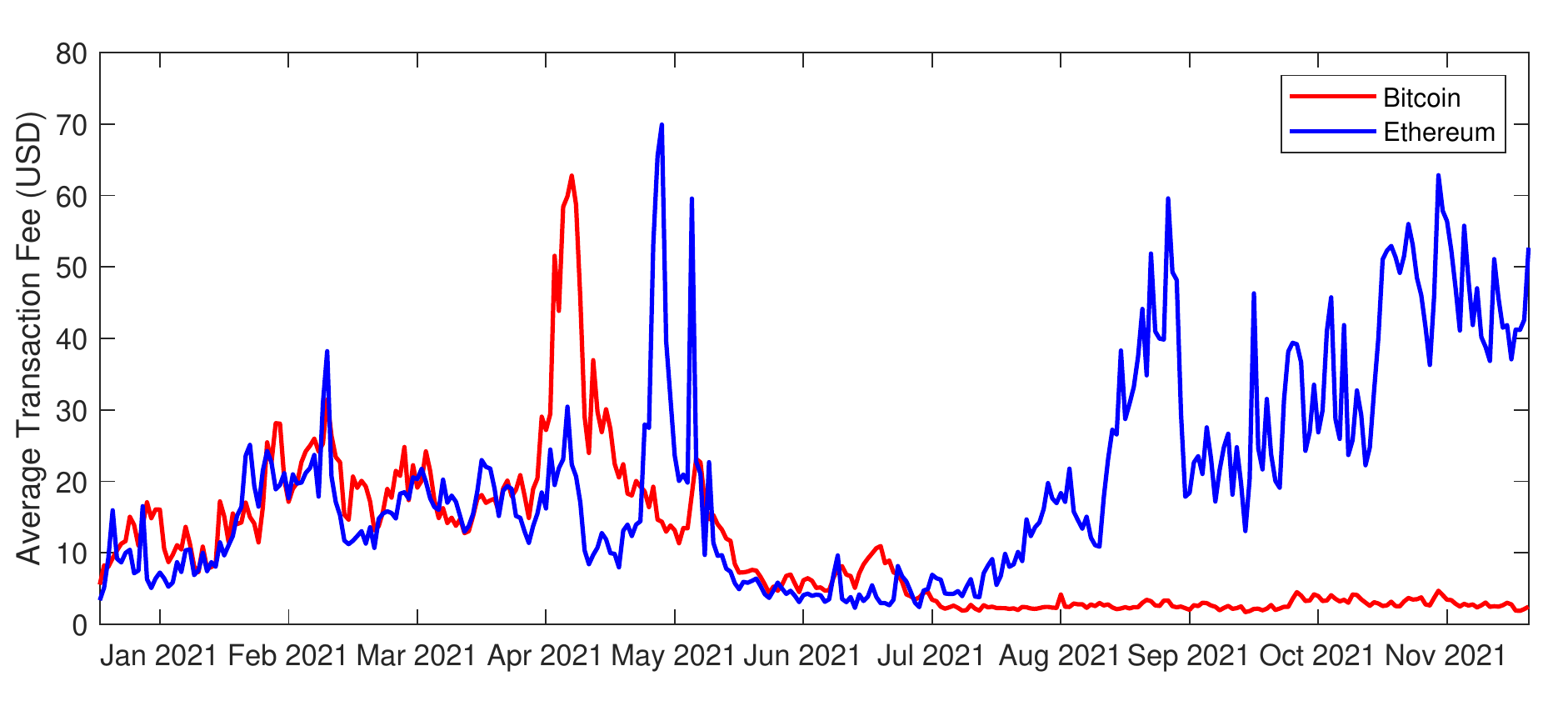}
	\caption{Average transaction fee for  Bitcoin and Ethereum blockchain platforms \cite{BEprice}.} \label{C2FIG3}	
\end{figure}

Third, the transaction costs of the current blockchain platforms are  high compared to the conventional IT based trading systems. Fig. \ref{C2FIG3} shows the average transaction fee for two of the most prominent blockchain platforms, i.e., Bitcoin and Ethereum. The high transaction cost would pose a barrier for prosumers to participate in the peer-to-peer energy trading.

\section{Artificial intelligence supporting operations of power systems}\label{S4}
The advanced metering infrastructures of smart grids produce a substantial volume of useful data. This data could be exploited by the AI to improve the situational awareness and operability of power systems.  This is particularly useful when small or medium sized prosumers participate in the operations of power systems and make decisions independently, given limited budgets of their control systems. According to the research \cite{dupont2012linear}, the cost of control solution at the household level is on the order of 57 USD per year. In this section, the state-of-the-art approaches of the AI applied in the operations of power systems are reviewed.

\subsection{Analysing and optimising operations of power systems}
This subsection introduces the approaches implemented to analyse and optimise the operations of power systems. The game theory is a collection of analytical tools for modelling the strategic decision making and interactions among stakeholders in power systems. The optimisation provides a solution to find optimal decisions for delivering certain objectives. Additionally, uncertainties caused by the intermittency of RESs and flexible demand would affect the accuracy of the power systems modelling. This review also discusses the statistical approaches for predicting these uncertainties.

\subsubsection{Game theory}
The Cournot and Stackelberg are two classic models for analysing decision making of stakeholders. The Cournot model describes that market players supply homogeneous products, and compete on the amount of supplied products by making decisions independently and simultaneously \cite{varian2014intermediate}. The Stackelberg model features a hierarchical two-level or multi-level sequential decision making process \cite{von2010market}. For the two-level decision making, the players are categorised into a leader-level which makes decisions first and a follower-level which makes subsequent decisions  responding to the leader's strategies. For the multi-level decision making, after the first level of followers makes responding decisions, they become a leader-level to make decisions prioritising the decisions of the next level of followers. This process continues until the last level of followers makes their responding decisions.

\subsubsection{Optimisation approaches}
The optimisation approaches can be categorised as programming techniques and heuristic algorithms. The programming techniques include the linear programming, integer linear programming, mixed integer linear programming, and non-linear programming. The linear programming refers to an optimisation problem in which all objective functions and constraints are the linear functions of decision variables \cite{dantzig1998linear}. In the integer linear programming, only binary values and integers can be used as decision variables \cite{schrijver1998theory}. In the mixed integer linear programming, both integers and non-integers can be used as decision variables \cite{milp}. The non-linear programming refers to an optimisation problem in which at least one  objective function or constraint is the non-linear function of decision variables \cite{bazaraa2013nonlinear}.

Although the non-linear programming problems can accurately model the practical operations of power systems, it is difficult to be solved by analytical approaches and ensure the global optimality. To overcome this issue, further research has focused on the heuristic algorithms. The heuristic algorithms can iteratively search over the entire feasible space to guarantee the global optimal solutions, including the particle swarm algorithm, genetic algorithm, artificial immune algorithm, and other heuristic algorithms. The particle swarm algorithm \cite{shi2001particle} optimises a problem by searching from solution set consisting of particles, and moving particles within the searching space according to predefined functions of particles' position and velocity. The movements of particles are determined by both the local best known position and global best known position of the searching space. A swarm of particles would ultimately move towards the best solution. The genetic algorithm \cite{deb2002fast} is based on the Darwin's theory of evolution, by which a population of candidate solutions to an optimisation problem is randomly generated and defined as a generation. The values of objective functions for every individual in the population are evaluated and defined as the fitness. The highly fitted individuals are selected and mutated to form a new generation. The population is iteratively evolved towards the best solution. Analogous to the genetic algorithm, for the artificial immune algorithm \cite{dasgupta2012artificial}, a population of candidate solutions to an optimisation problem is randomly generated and defined as antigens. The values of objective functions for every antigen in the population are evaluated and defined as the antibodies. The antigens are iteratively cloned towards the best solution.

\subsubsection{Research and implementations}
The game theory and optimisation have been well documented by research on the power systems scheduling. The game-theoretic models, players, solving approaches, advantages, challenges, differences, and commons from the current research are summarised in Table \ref{TABLE21}. Belgana \emph{et al.} \cite{belgana2014open} developed a multi-leader and multi-follower Stackelberg game-theoretic problem to find optimal strategies that could maximise profits of utilities and minimise carbon emissions. The problem was solved by a hybrid multiobjective evolutionary algorithm. The results demonstrated a trade-off between emissions, profits, and bills, while there was an opportunity to improve the searching mechanism of the algorithm and consider power losses. Meng and Zeng \cite{meng2013stackelberg} proposed a 1-leader, n-follower  Stackelberg game to maximise the profits of retailers at the leader-level and minimise the electricity bills of consumers at the follower-level considering the real-time pricing scheme. The genetic algorithm was used to solve the leader's optimisation problem and the linear programming was used to solve the follower's optimisation problem. This  Stackelberg game yielded an efficient retail pricing to incentivise the demand response of consumers, whereas the competitive retail markets and imperfect information from consumers could be extended by this model. Analogously, a Stackelberg game-theoretic problem was proposed in \cite{9016641} to model the interactions between the policy maker and generators/consumers for decarbonising power systems, which was solved by the designed bi-level multiobjective immune algorithm. This  model could reduce carbon emissions and improve social welfare for the GB energy system. However, how to reallocate carbon revenue received from energy sector remained as a challenge. Ghosh \emph{et al.}  \cite{ghosh2018exchange} formulated a coupled constrained potential game to set the energy exchange prices for  maximising the amount of the energy exchange among prosumers and reducing the consumption from the utility grid.  A distributed algorithm was proposed enabling individual prosumers to optimise their own payoffs. This work could be further extended to consider the non-linear price of prosumers and uncertainties caused by RESs. The Cournot game was implemented in \cite{8882303} to model the competition between customers and utilities in distribution networks for satisfying  the system reliability. How to facilitate the engagement of consumers into the reliability improvement remained to be considered. Similarly, Zhang \emph{et al.}\cite{vuelvas2019novel}  modelled the local energy trading as a non-cooperative Cournot game to stimulate the regional energy balance and promote the penetration of RESs, while the transmission and intertemporal constraints could be considered to improve the accuracy of the proposed model.

\begin{table}[H]\scriptsize
	\caption{Comparison of literatures on game-theoretic approaches in the field of power systems operation.}
	\centering
	\label{TABLE21}
	\begin{tabular}{lllllll}
		\hline	
		\!\!\!\!\!\! Literature   &\!\!\!\!\!\!\!\!\! Belgana \emph{et al.} \cite{belgana2014open}&\!\!\!\!\!\!\!\!\! Meng and Zeng  \cite{meng2013stackelberg}&\!\!\!\!\!\!\!\!\! Hua \emph{et al.} \cite{9016641}&\!\!\!\!\!\!\!\!\! Ghosh \emph{et al.} \cite{ghosh2018exchange}&\!\!\!\!\!\!\!\!\! Mohammadi \emph{et al.} \cite{8882303}&\!\!\!\!\!\!\!\!\! Zhang \emph{et al.} \cite{vuelvas2019novel} \\	\hline			
				
		\!\!\!\!\!\!  Model  &\!\!\!\!\!\!\!\!\! Stackelberg&\!\!\!\!\!\!\!\!\! Stackelberg&\!\!\!\!\!\!\!\!\! Stackelberg	&\!\!\!\!\!\!\!\!\! Potential game&\!\!\!\!\!\!\!\!\! Cournot &\!\!\!\!\!\!\!\!\! Cournot\\
        \!\!\!\!\!\! Player&\!\!\!\!\!\!\!\!\! Microproducers and &\!\!\!\!\!\!\!\!\! Retailer and consumers&\!\!\!\!\!\!\!\!\! Policy maker, generators&\!\!\!\!\!\!\!\!\! Utility and prosumers	&\!\!\!\!\!\!\!\!\! Customers and utilities	&\!\!\!\!\!\!\!\!\! Energy providers\\
         &\!\!\!\!\!\!\!\!\! consumers &\!\!\!\!\!\!\!\!\! &\!\!\!\!\!\!\!\!\! and consumers	&\!\!\!\!\!\! \!\!\!	&\!\!\!\!\!\! \!\!\!	&\!\!\!\!\!\!\!\!\!  \\            
       \!\!\!\!\!\!  Solution &\!\!\!\!\!\!\!\!\! Hybrid multiobjective &\!\!\!\!\!\!\!\!\! Genetic algorithm  and &\!\!\!\!\!\!\!\!\! Bi-level multiobjective  &\!\!\!\!\!\!\!\!\! Distributed algorithm &\!\!\!\!\!\!\!\!\! Lagrangian function 	&\!\!\!\!\!\!\!\!\! Optimal generation\\ 	
        &\!\!\!\!\!\!\!\!\! evolutionary algorithm 	& \!\!\!\!\!\!\!\!\!  linear programming      &  \!\!\!\!\!\!\!\!\!         immune algorithm &\!\!\!\!\!\!\!\!\! Distributed algorithm &\!\!\!\!\!\!\!\!\! and KKT conditions 	&\!\!\!\!\!\!\!\!\!  plan algorithm\\
        
     \!\!\!\!\!\! Advantage &\!\!\!\!\!\!\!\!\! Trade-off between	          & \!\!\!\!\!\!\!\!\! Efficient retail pricing  & \!\!\!\!\!\!\!\!\! Carbon reduction and    &\!\!\!\!\!\!\!\!\! Local energy balance  &\!\!\!\!\!\!\!\!\! Customer reliability	&\!\!\!\!\!\!\!\!\! Peak shifting  \\        
             \!\!\!\!\!\!  &\!\!\!\!\!\!\!\!\! emissions,	profits, and bills & \!\!\!\!\!\!\!\!\!                          & \!\!\!\!\!\!\!\!\!  improving social welfare  &\!\!\!\!\!\!\!\!\! and Peak reduction  &\!\!\!\!\!\!\!\!\! 	&\!\!\!\!\!\!\!\!\!  \\

     \!\!\!\!\!\! Challenges &\!\!\!\!\!\!\!\!\! Improving searching	& \!\!\!\!\!\!\!\!\! Considering competitive  & \!\!\!\!\!\!\!\!\! Addressing carbon  &\!\!\!\!\!\!\!\!\! Considering non-linear  &\!\!\!\!\!\!\!\!\! Facilitating consumers'	&\!\!\!\!\!\!\!\!\! Including transmission  \\
     \!\!\!\!\!\!  &\!\!\!\!\!\!\!\!\! mechanism  and considering	& \!\!\!\!\!\!\!\!\! market and imperfect  & \!\!\!\!\!\!\!\!\! revenue reallocation   &\!\!\!\!\!\!\!\!\! pricing and uncertainty  &\!\!\!\!\!\!\!\!\! engagement	&\!\!\!\!\!\!\!\!\! and intertemporal  \\ 
 
      \!\!\!\!\!\!  &\!\!\!\!\!\!\!\!\!  power losses	& \!\!\!\!\!\!\!\!\!  information  & \!\!\!\!\!\!\!\!\!  &\!\!\!\!\!\!\!\!\!   &\!\!\!\!\!\!\!\!\! 	&\!\!\!\!\!\!\!\!\! constraints \\

     \!\!\!\!\!\! Context &\!\!\!\!\!\!\!\!\! Microgrids	& \!\!\!\!\!\!\!\!\! Retail market  & \!\!\!\!\!\!\!\!\! Whole power system  &\!\!\!\!\!\!\!\!\! Distribution network  &\!\!\!\!\!\!\!\!\! Distribution network	&\!\!\!\!\!\!\!\!\! Whole power system  \\
         \!\!\!\!\!\! difference&\!\!\!\!\!\!\!\!\!	& \!\!\!\!\!\!\!\!\!  & \!\!\!\!\!\!\!\!\!   &\!\!\!\!\!\!\!\!\!  &\!\!\!\!\!\!\!\!\!	&\!\!\!\!\!\!\!\!\!  \\     	
            \!\!\!\!\!\! Common&\multicolumn{6}{l}{\!\!\!\!\!\!\!\!\! Implementing the game theory  for modelling and analysing the decision making and interactions of stakeholders in the operation of power systems} \\   
        
		\hline			
	\end{tabular}
\end{table}

\subsubsection{Analysis of uncertainties in power systems}
It is crucial for individual prosumers and power systems to account possible variations of uncertainties when modelling and optimising the operations.  Using a set of scenarios is a statistical approach to predict these variations, by which each variation is defined as a scenario \cite{morales2010methodology}. The uncertain scenarios are generated from the probabilistic  distributions of historical data by using sampling approaches \cite{hasan2019existing}, such as the Monte Carlo simulation  \cite{santos2016methodology, hemmati2019uncertainty}, Latin hypercube sampling \cite{preece2015efficient, 7285758, xiao2018application} and stochastic analysis \cite{mavromatidis2018design, 7439085}. The typical literatures, uncertain variables, advantages, challenges, differences, and commons  are summarised in Table \ref{TABLE24}. Santos \emph{et al.}  \cite{santos2016methodology} implemented the Monte Carlo simulation to generate scenarios of RESs and solved the system optimisation problem under these scenarios using the deterministic approach. The proposed approach improved the computational efficiency compared to stochastic approaches, while the responding measures to the predicted uncertainties could be considered into the optimisation problem.  Similarly, Hemmati \emph{et al.} \cite{hemmati2019uncertainty} analysed  the uncertainties of RESs and load deviation by the Monte Carlo simulation, and  incorporated the uncertainty analysis into the decision making process to maximise the generating profits.  The proposed approach helped system operators make better decisions on the energy dispatch under these uncertainties. However, the computational efficiency of the proposed approach could be further improved.

The  Monte Carlo simulation would cause the issues of computationally intensive and inefficiency due to the high standard deviations of samples caused by the randomly sampling. These issues can be overcome by the Latin Hypercube sampling, since the space-filling of the Latin Hypercube sampling would reduce the standard deviation of samples. In \cite{preece2015efficient}, the Latin hypercube sampling was used to generate uncertain scenarios of intermittent generation  for overcoming those issues of the Monte Carlo simulation and considered the low-probable conditions. The results yielded accurate predictions for uncertainties with small samples, whereas there was an opportunity to consider the online probabilistic analysis.  Mavromatidis \emph{et al.} \cite{mavromatidis2018design} proposed a two-stage stochastic programming combined with the Latin Hypercube sampling to incorporate the uncertainties of the energy prices, emissions factors, heating demand, electricity demand, and solar radiation into the scheduling of distribution systems. This study demonstrated that the designed stochastic method could yield a more accurate estimation of these uncertainties than deterministic methods,  while decision criteria representing risk levels could be considered into the decision making process. Huang \emph{et al.} \cite{7439085} designed an economic dispatch model for virtual power plants, by which the uncertainties caused by the RESs and flexible demand were described  by the stochastic intervals. These intervals were subsequently integrated into the problem of minimising costs. The results demonstrated the industrial applicability of the proposed approach without the need to obtain the probability density function. However, determining the probability degree for intervals remained a challenge.

Further research efforts have been dedicated to improving the predicting accuracy and adaptability of scenarios. Liang \emph{et al.} \cite{7285758} proposed a non-parametric kernel density estimation method to yield the probability density distribution of uncertainties from the behaviours of electric vehicles. The scenarios were generated from the probability density distribution through using the Latin hypercube sampling. However, the probability behaviours of mixed energy patterns of electric vehicles could be further investigated. To select the high-probable scenarios,  Xiao \emph{et al.} \cite{xiao2018application} proposed an statistical approach to merge scenarios with a minimum probability distance. The proposed approach selected typical scenarios for accurate prediction of uncertainties, while the scalability to other power networks remained a challenge.

\begin{table}[H]\scriptsize
	\caption{Comparison of literatures on scenarios approaches  in the field of power systems operation.}
	\centering
	\label{TABLE24}
	\begin{tabular}{lllllllll}
		\hline	
	 \!\!\!\!\!\! 	Literature   &\!\!\!\!\!\!\!\!\!\!\!  Santos \emph{et al.}  \cite{santos2016methodology}&\!\!\!\!\!\!\!\!\! \!\!	Hemmati \emph{et al.} \cite{hemmati2019uncertainty}&\!\!\!\!\!\!\!\!\!\!\! Preece \emph{et al.} \cite{preece2015efficient}&\!\!\!\!\!\!\!\!\!\!\! Mavromatidis \emph{et al.} \cite{mavromatidis2018design}&\!\!\!\!\!\!\!\!\!\!\! Huang \emph{et al.} \cite{7439085}&\!\!\!\!\!\!\!\!\!\!\! Liang \emph{et al.} \cite{7285758}&\!\!\!\!\!\!\!\!\!\!\! Xiao \emph{et al.}    \cite{xiao2018application}\\\hline	
	 \!\!\!\!\!\! 	Uncertain &\!\!\!\!\!\!\!\!\!\!\! RESs&\!\!\!\!\!\!\!\!\!\!\! RESs and load&\!\!\!\!\!\!\!\!\!\!\! Intermittent &\!\!\!\!\!\!\!\!\!\!\! Energy prices, carbon factors&\!\!\!\!\!\!\!\!\!\!\! RESs and load&\!\!\!\!\!\!\!\!\!\!\! Electric vehicle&\!\!\!\!\!\!\!\!\!\!\! RESs and load\\
	  \!\!\!\!\!\!     variables               	&\!\!\!\!\!\!\!\!\!\!\!    &  \!\!\!\!\!\!\!\!\!\!\!            &\!\!\!\!\!\!\!\!\!\!\! generation                       &\!\!\!\!\!\!\!\!\!\!\! demand, and solar radiation&\!\!\!\!\!\!\!\!\!\!\! &\!\!\!\!\!\!\!\!\!\!\! behaviours&\!\!\!\!\!\!\!\!\!\!\! \\
	  
	  \!\!\!\!\!\!    Advantage &\!\!\!\!\!\!\!\!\!\!\!  More efficient than &\!\!\!\!\!\!\!\!\!\!\! Improved optimal &\!\!\!\!\!\!\!\!\!\!\!  Accurate prediction &\!\!\!\!\!\!\!\!\!\!\! Accurate prediction&\!\!\!\!\!\!\!\!\!\!\!  Easy for&\!\!\!\!\!\!\!\!\!\!\! Without pre-defined &\!\!\!\!\!\!\!\!\!\!\! Remain typical\\

	  \!\!\!\!\!\!                     &\!\!\!\!\!\!\!\!\!\!\! stochastic approaches &\!\!\!\!\!\!\!\!\!\!\! energy dispatch &\!\!\!\!\!\!\!\!\!\!\!  with small samples &\!\!\!\!\!\!\!\!\!\!\! &\!\!\!\!\!\!\!\!\!\!\!  application&\!\!\!\!\!\!\!\!\!\!\! density functions&\!\!\!\!\!\!\!\!\!\!\! scenarios\\

	  \!\!\!\!\!\!    Challenge &\!\!\!\!\!\!\!\!\!\!\!  Including responding &\!\!\!\!\!\!\!\!\!\!\! Reducing computing &\!\!\!\!\!\!\!\!\!\!\!  Considering online &\!\!\!\!\!\!\!\!\!\!\! Including decision&\!\!\!\!\!\!\!\!\!\!\! Determining&\!\!\!\!\!\!\!\!\!\!\! Behaviours of mixed&\!\!\!\!\!\!\!\!\!\!\! Scalability\\	  
	
	  \!\!\!\!\!\!    &\!\!\!\!\!\!\!\!\!\!\!  measures &\!\!\!\!\!\!\!\!\!\!\! time                                    &\!\!\!\!\!\!\!\!\!\!\!  probabilistic analysis &\!\!\!\!\!\!\!\!\!\!\!  criteria&\!\!\!\!\!\!\!\!\!\!\! probability degree&\!\!\!\!\!\!\!\!\!\!\! electric vehicles&\!\!\!\!\!\!\!\!\!\!\!  \\	  
	  
	  \!\!\!\!\!\!   Approach &\!\!\!\!\!\!\!\!\!\!\! Monte Carlo &\!\!\!\!\!\!\!\!\!\!\! Monte Carlo &\!\!\!\!\!\!\!\!\!\!\!  Latin Hypercube &\!\!\!\!\!\!\!\!\!\!\! Latin Hypercube&\!\!\!\!\!\!\!\!\!\!\! Stochastic intervals&\!\!\!\!\!\!\!\!\!\!\! Latin Hypercube&\!\!\!\!\!\!\!\!\!\!\! Latin Hypercube\\
	    
	  \!\!\!\!\!\!    difference&\!\!\!\!\!\!\!\!\!\!\! simulation&\!\!\!\!\!\!\!\!\!\!\! simulation &\!\!\!\!\!\!\!\!\!\!\!  sampling&\!\!\!\!\!\!\!\!\!\!\!  sampling&\!\!\!\!\!\!\!\!\!        &\!\!\!\!\!\!\!\!\!\!\!  sampling&\!\!\!\!\!\!\!\!\!\!\!   sampling\\          		
		   \!\!\!\!\!\! Common &\multicolumn{7}{l}{\!\!\!\!\!\!\!\!\!\!\! Using statistical approaches to generate scenarios for predicting potential variations of uncertain variables} \\  
            
		\hline
	\end{tabular}
\end{table}

\subsection{Data-driven machine learning}
The machine learning is capable of exploiting historical data  to capture typical features of actors in the operation of power systems, and improving the scalability and computational efficiency from using optimisation approaches.

\subsubsection{Learning approaches}
The learning approaches can be categorised as the supervised learning, unsupervised learning, and reinforcement learning. For the supervised learning, the input is provided as a labelled dataset, such that the model can learn from the labels to improve the learning accuracy \cite{jordan1992forward}. By contrast, for the unsupervised learning, there is no labelled dataset, such that the model explores the hidden features and predicts the output in a self-organising manner \cite{barlow1989unsupervised}. For the reinforcement learning, the model learns to react to the environment by self-adjusting through travelling from one state to another \cite{sutton2018reinforcement}.

\subsubsection{Research and implementations}
Applying learning approaches in solving decision making problems during the operations of power systems has been well studied in literatures. The typical literatures, targeted issues, advantages, challenges, differences, and commons  are summarised in Table \ref{TABLE23}. Zhang \emph{et al.} \cite{8886404} developed an online learning approach to replace heuristic algorithms for solving a cost minimisation problem under the uncertain RESs and loads.  The results demonstrated the improved solution optimality and computational efficiency compared to heuristic algorithms, while the location planning of RESs could be considered into this problem. Mbuwir \emph{et al.} \cite{9016682} compared two approaches of the reinforcement learning, i.e., the policy iteration and  fitted Q-iteration, in terms of scheduling the operation of the battery and heat pump in a residential microgrid. The simulation results demonstrated that the  policy iteration outperformed the fitted Q-iteration, and both approaches outperformed the optimisation approach in terms of improving the computational efficiency, whereas the future work could be extended to consider the grid congestions and energy sharing. In \cite{8918214}, a Q-learning algorithm was used as a reinforcement learning approach to minimise the costs and protect the privacy  when the EV owners exchange energy, while the physical constraints for the vehicle-to-grid (V2G) could be further considered.  Analogously, Shafie-Khah \emph{et al.} \cite{9250226} designed a Q-learning algorithm to optimally submit the bids of demand response for the end-user. The numerical results proved that the proposed model could reduce the costs of using electricity and improve the load balance. However, how to facilitate the participation of consumers to the scheme of  demand response could be considered. An energy management system was designed in \cite{7042790} to provide demand response services, by which the predefined model of consumers' dissatisfaction was replaced by the feature representations extracted through using the reinforcement learning. The designed energy management system enabled flexible requests from different users. However, how to avoid a new peak caused by the synchronised demand response of a large number of consumers remained a challenge. Ruelens \emph{et al.} \cite{7038106} combined the heuristic algorithm with reinforcement learning to control a cluster of loads and storage devices. The simulation demonstrated the effectiveness on the cost reduction through using the proposed algorithm, while the exploration strategies could be included to improve the learning efficiency. Gasse \emph{et al.} \cite{gasse2019exact} proposed a learning model for extracting the branch-and-bound variable selection policies to solve the combinatorial optimisation problem, and testified that a series of computational complex problems could be efficiently solved. How to apply the proposed learning model for assisting other heuristic algorithms could be further explored. Zhang \emph{et al.} \cite{7450180} integrated the learning mechanism with optimisation techniques to obtain optimal demand response policies. The designed controller could help consumers reduce electricity bills with improved computational efficiency. However, how to capture the load change between two set-points remained a challenge.
\begin{table}[H]\scriptsize
	\caption{Comparison of literatures on learning approaches  in the field of power systems operation.}
	\centering
	\label{TABLE23}
	\begin{tabular}{llllll}
		\hline	
		Literature   &  Zhang \emph{et al.} \cite{8886404}&Mbuwir \emph{et al.} \cite{9016682}&Najafi \emph{et al.}  \cite{8918214}&Shafie-Khah \emph{et al.} \cite{9250226}\\	\hline	
		Targeted issue&Replacing heuristic algorithms &Extracting policy function and Q-function & Extracting Q-function& Extracting Q-function\\
		Advantage& Improved optimality and&Improved computational efficiency&Protected privacy and &Cost minimisation and\\
			     &  computational efficiency      &                                & cost minimisation&load balance\\
		Challenge&Considering locational planning  &Solving grid congestions and energy &Considering physical &Facilitating consumers'\\
	             &                                 &sharing                         & constraints of V2G&participation\\

		Approach  &	Online convex optimization&	Reinforcement learning&	Reinforcement learning&	Reinforcement learning\\
			difference  & &&&\\
		
         \hline	
		
        Literature   & Wen  \emph{et al.}   \cite{7042790} &Ruelens \emph{et al.} \cite{7038106}&Gasse \emph{et al.} \cite{gasse2019exact}&Zhang \emph{et al.} \cite{7450180}\\ \hline	
            
        Targeted issue& Extracting feature representations  &Extracting policy function  &Extracting  policy function  &Extracting policy function\\

      	Advantage&Flexible request of users&   Cost reduction      &Computational efficient&Bill reduction\\
        Challenge&Avoiding synchronised demand&Including exploration strategies &Application on assisting other &Considering load change\\
                  &response & & heuristic algorithms &\\        
        
        Approach  & Reinforcement learning      &Heuristic algorithm and &     Graph convolutional & Optimisation and neural \\     
        difference   &    &reinforcement learning&     neural network &  network   \\       
        Common &\multicolumn{3}{l}{	Applying learning approaches in solving decision making problems during the operations of power systems}\\
        \hline
	\end{tabular}
\end{table}

\subsection{Remark of practicalities, benefits, and challenges}
This section remarks the extents for implementing the AI into operations of power systems, and outlines the benefits and potential challenges of such implementations. 
\subsubsection{Practical implementations}
\begin{figure}[H]
	\centering
	\includegraphics [width=4.5in]{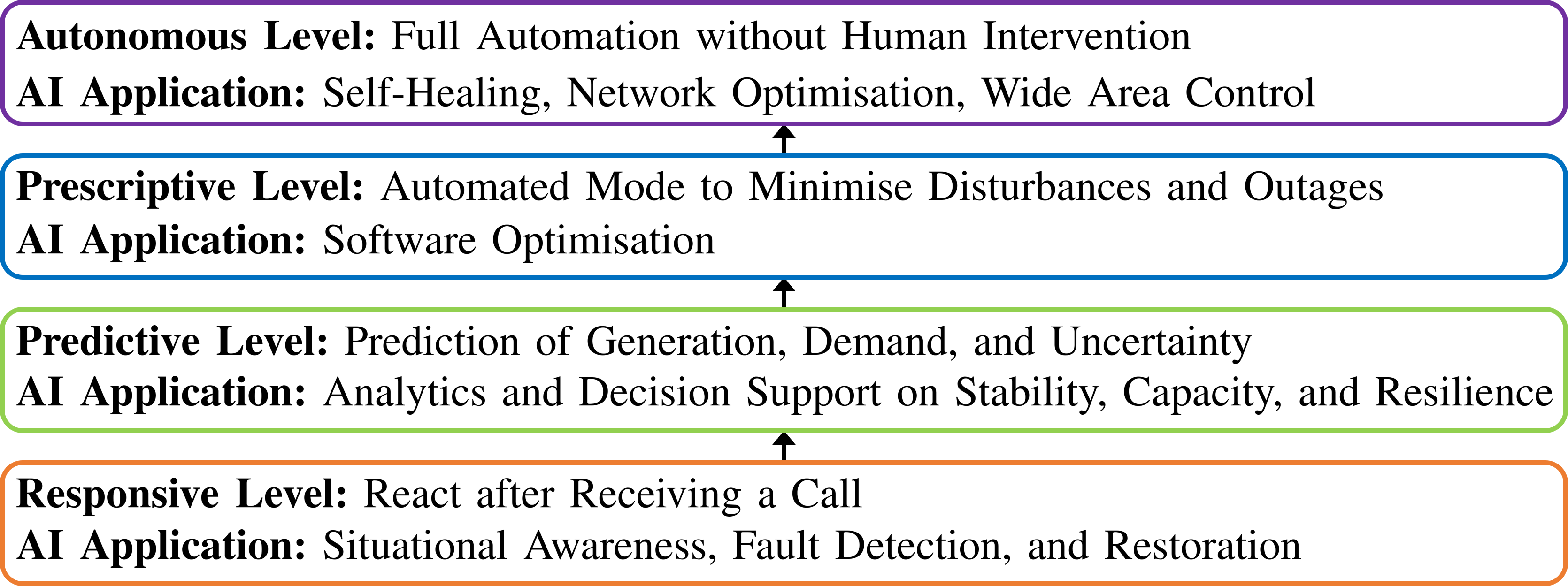}
	\caption{Four extents of implementing the artificial intelligence into the operations of power systems.} \label{C2FIG4}	
\end{figure}

The extents of implementing the AI into operations of power systems can be categorised into four levels as shown in Fig. \ref{C2FIG4}. The first level, i.e., responsive level, features the conventional operations of power systems, in which the analytical AI assists the situational awareness, fault detection, and restoration of power systems after receiving a call of outages. The development of the deep AI and digitalisation of power systems enable the transition towards the second level, i.e., predictive level, where more AI based analytics and decision supporting tools are included to predict the real-time generation, demand, and uncertainties, so as to maintain system performances, e.g., stability, capacity margin, and resilience. At the third level, i.e., prescriptive level, a number of functions in the first two levels can be performed automatically with the support of AI based software, so as to minimise the disturbances and outages from the systematic perspective. At the fourth level, i.e., autonomous level, full  automation of system operations would be achieved in the future, where the wide area controlling decisions and network optimisations could be delivered by an AI based digital layer without the intervention of system operators, so that the system can maintain the self-healing.

\subsubsection{Benefits}
The benefits of implementing the AI into the operations of power systems are summarised as follows:

$\bullet$ \emph{Automation:} The AI can automatically make optimal decisions for generators (e.g., generation dispatch), power system operators (e.g., state estimation, real-time control, energy balancing, and contingency screening), consumers (e.g., smart load control).

$\bullet$ \emph{Computational efficiency:} The AI can improve the computational efficiency for the management of power systems, so as to achieve the real-time reactive operations.

$\bullet$ \emph{Interoperability:} The AI can assist the strategic coordination of actors in power systems, in achieving the overall system benefits.

$\bullet$ \emph{Scalability:} The developed AI based models and software are scalable for different scales of power systems from households, industries, businesses,  to an entire region, since they only require the historical data to extract typical characteristics of these systems. 

$\bullet$ \emph{Adaptability:} The AI based management tools can dynamically adapt the situations of power systems, which ensures the system resilience under both the high-probable but low-impact issues, e.g., power imbalance, and low-probable but high-impact issues, e.g., extreme weather conditions.

\subsubsection{Challenges}
There are two primary issues when implementing the AI to the operations of power systems: 

First, when the historical data is too less to train an accurate AI model, how to guarantee the model accuracy and avoid the issues of over/under fitting presents a challenge. 

Second, how to ensure that controlling decisions yielded by the AI models align with the physical constraints of power systems presents another challenge.

\section{Conclusion}\label{S5}
To investigate how to exploit the blockchain and AI for facilitating the emerging role of prosumers to be integrated into smart grids and decarbonising the power systems, a comprehensive review from the aspects of the regulations, energy markets, and operations of power systems is provided by this paper. This review particularly focuses on the state-of-the-art research and applications of the blockchain and AI in terms of supporting the decentralised energy trading and decision making during the operations.
From the regulatory perspective, the vital barrier for facilitating the engagement of prosumers is the lack of dynamic and decentralised policy measures. Overcoming this barrier requires future research and practical regulatory design to identify 
key responsibilities, assets, roles, and  models for prosumers. From the market perspective, the vital issue  for accommodating the new role of prosumers is to design appropriate local market structures so as to align individual profits with system benefits, which requires the focus on the rulesets, pricing, transactions, trading platforms, and auction mechanisms. From the operational perspective, the vital issue is to fit novel AI models into the physical operations and constraints of power systems. This requires the transition towards a digitalised and interoperable power systems with intensive interactions between the digital layer and physical layer. Therefore, this review concludes that by incorporating the blockchain and AI, the smart grids 
can support the integration of prosumers with the functions of trading, control, and policy. Nonetheless, this is achievable only if the vital issues and barriers on the regulation, market, and operation are overcome.

\bibliography{review}
\bibliographystyle{elsarticle-num} 

\end{document}